\documentclass[fleqn,twoside,twocolumn,nofootinbib,showkeys]{revtex4} 
\usepackage[sec,nocpr]{ujp} 

\begin{document}
\title[Inefficiency of $\delta$-interaction Potentials in Multi-Dimensional Spaces]
{ON INEFFICIENCY OF REPULSIVE\\ \boldmath$\delta$-POTENTIALS  IN
MULTIDIMENSIONAL SPACES}
\author{I.V.~Simenog}
\affiliation{Bogolyubov Institute for Theoretical Physics, Nat.
Acad. of Sci. of Ukraine}
\address{14-b, Metrolohichna Str., Kyiv, 03680, Ukraine}
\email{bgrinyuk@bitp.kiev.ua}
\author{B.E.~Grinyuk}%
\affiliation{Bogolyubov Institute for Theoretical Physics, Nat.
Acad. of Sci. of Ukraine}%
\address{14-b, Metrolohichna Str., Kyiv, 03680, Ukraine}%
\author{M.V.~Kuzmenko}
\affiliation{Bogolyubov Institute for Theoretical Physics, Nat.
Acad. of Sci. of Ukraine}
\address{14-b, Metrolohichna Str., Kyiv, 03680, Ukraine}
\email{bgrinyuk@bitp.kiev.ua}  \udk{530.145} 
\razd{\secix}

\autorcol{I.V.\hspace*{0.7mm}Simenog, B.E.\hspace*{0.7mm}Grinyuk,
M.V.\hspace*{0.7mm}Kuzmenko}

\setcounter{page}{1}%

\begin{abstract}
A complete account of correlations has been shown to make
$\delta$-like repulsive interaction potentials inefficient for any
$N$-particle quantum system in the $D$-dimensional space with
$D\geq2$.
\end{abstract}
 \maketitle

\section{Introduction}

Since the works by Fermi and G\"{o}ppert-Mayer, simplified versions
of potentials in the form of $\delta$-like interaction potentials
have been used in various fields of theoretical physics. In one
time, a discussion of the role of relativistic correction terms in
the form of $\delta$-potentials to the Coulomb interaction in the
QED was problematic when considering the quasi-relativistic Breit
potentials \cite{R1}. Modern studies of Bose condensation effects
are based on using the self-consistent Gross--Pitaevskii field
\cite{R2, R3} and the idea of $\delta$-like interaction potentials
(see works \cite{R4, R5}). In general, various $D$-dimensional
nonlinear evolution equations like the nonlinear Schr\"{o}dinger one
are often considered as associated with certain many-body systems
characterized by $\delta$-like interaction potentials. Some versions
of effective Skirm forces \cite{R6} (see also works \cite{R7, R8})
in the form of superposition of two- and three-particle
$\delta$-like potentials remain still popular in nuclear physics as
a model of interaction between nucleons, which is used to describe
the structure characteristics of atomic nuclei, from light and
intermediate ones up to the most heavy nuclei, within the simplest
one-particle self-consistent mean field approximation.

It should be noted that a consideration of the problems containing $\delta
$-potentials is very often restricted to calculations in the first-order
approximation of perturbation theory (which are convenient to be done just
with such potentials) and assuming that the exact solution of the problem does
not change the result from the qualitative point of view. However, such an
assumption has not been substantiated. In this work, we study the issue of
whether $\delta$-potentials are applicable and effective in the case of
many-body systems in a $D$-dimensional space, as well as the role of the
consistent account of pair correlation effects in such problems.

\section{Formulation\\ and Preliminary Analysis of the Problem}

Consider a quantum-mechanical system of $N$ particles in a
$D$-dimensional space. In addition to some usual potentials,
$U\left(  r_{ij}\right)  $, the interaction between particles also
includes $\delta$-like potentials. Moreover, the system can be
located in an external potential field $V\left( \mathbf{r}\right) $.
As a result, the system Hamiltonian looks like\vspace*{-1mm}
\[
\hat{H} = \sum\limits_{i=1}^{N} \left(\!
\frac{\mathbf{p}_{i}^{2}}{2m}+V\left( \mathbf{r}_{i} \right)
\!\right)+\sum\limits_{j>i=1}^{N}U\left(r_{ij}\right)+
\]\vspace*{-5mm}
\begin{equation}
+\,
g\sum\limits_{j>i=1}^{N}\delta_{\varepsilon}\left(\mathbf{r}_{ij}\right)\!.
\label{E1}
\end{equation}%
Hereafter, the $\delta$-functions are defined by means of a sequence
of $\delta_{\varepsilon}$-like functions, in particular, in the
form\vspace*{-1mm}
\begin{equation}\label{E2}
\delta_{\varepsilon}\left( \mathbf{x}
\right)=\frac{1}{\left(\sqrt{\pi}\,\varepsilon\right)^{D}}\,
e^{-\mathbf{x}^2/\varepsilon^2}, \quad \delta_{\varepsilon}\left(
\mathbf{x} \right)\,\,\overrightarrow{_{_{_{_{\varepsilon\rightarrow
0 }}}}}\,\, \delta\left(\mathbf{x}\right)\!,
\end{equation}
where $\mathbf{x}^{2}\equiv\sum_{k=1}^{D}x_{k}^{2}$ is the squared
interval in the $D$-dimensional space. Generally speaking, the
specific profile of $\delta_{\varepsilon}$ is not too important, and
it is chosen in form (\ref{E2}) for convenience. It is important
that the limit $\varepsilon \rightarrow0$ should be understood as
the one carried out in the final solutions obtained for a given
$\varepsilon$.

Let us first consider, for simplicity, a model consisting of two particles
interacting via the oscillator and repulsive $\delta$-like potentials (in the
center of mass frame, with the unit reduced mass and the unit circular
oscillation frequency):
\begin{equation}\label{E3}
\left(\!-\frac{1}{2}\triangle+\frac{1}{2}\mathbf{r}^{2}+g\delta\left(
\mathbf{r}\right)\!\right)
\psi\left(\mathbf{r}\right)=E\psi\left(\mathbf{r}\right)\!.
\end{equation}
What physical consequences follow from the availability of
$\delta$-potential in Eq.~(\ref{E3}) in the $D$-dimensional space?
In the one-dimensional case, one can find that, due to the repulsive
$\delta$-potential, all even-parity oscillator levels would shift
upwards, and as $g\rightarrow\infty$, they would approach the
neighbor odd-parity oscillator levels. At the same time, the
odd-parity oscillator levels would not be shifted by the
$\delta$-potential, which turns out inefficient for them (because
the corresponding wave functions equal zero just at the point where
the $\delta$-function is located).

Now let us consider a nontrivial three-dimensional case for Eq.~(\ref{E3}) and
a spherically symmetric state of the system (since the $\delta$-potential is
not efficient for the states with non-zero angular momenta owing to the factor
$\sim r^{l}$ in the wave function). Expanding the solution of Schr\"{o}dinger
equation (\ref{E3}) into a series of oscillator eigen-functions for the
zero-order (unperturbed by the $\delta$-potential) problem,
\begin{equation}
\psi\left(  r\right)  =\sum\limits_{k}c_{k}\psi_{k}\left( r\right)\!
,\label{E4}%
\end{equation}
we obtain the explicit solution
\begin{equation}\label{E5}
\psi\left(r\right)=\frac{1}{\sqrt{\sum\limits_{i=0}^{K}
\frac{\left|\psi_{i}\left(0\right)\right|^{2}}{\left(E-E_{i}\right)^{2}}}}
\sum\limits_{k=0}^{K}\frac{\psi_{k}^{\ast}\left(0\right)}
{E-E_{k}}\psi_{k}\left(r\right)\!,
\end{equation}
which approaches the exact one as $K$ grows. At
$K\rightarrow\infty$, series (\ref{E5}) would have been an exact
solution if it has been convergent. The energy levels in
$D$-dimensional problem (\ref{E5}) are determined from the secular
equation
\[
\frac{1}{g}=\sum\limits_{k=0}^{K}\frac{\left|\psi_{k}\left(0\right)\right|
^{2}}{E-E_{k}}=
\]\vspace*{-7mm}
\begin{equation}\label{E6}
=\frac{1}{\pi^{D/2}\Gamma\left(D/2\right)}
\sum\limits_{k=0}^{K}\frac{1}{\Delta-2k}
\frac{\Gamma\left(k+D/2\right)}{\Gamma\left(k+1\right)},
\end{equation}
where $\Delta\equiv E-D/2$\thinspace\ is the energy shift, and
$\Gamma\left( z\right)  $  is the Euler gamma function. At a fixed
$K$, transcendental equation (\ref{E6}) has a solution $\Delta_{0}$
describing the upward shift of the ground state energy, which
obviously falls within the interval $0<\Delta_{0}<2$. The shift
$\Delta_{1}$ of the first excited state lies within the interval
$2<\Delta_{1}<4$; the shift of the next level, $\Delta _{2}$, within
the interval $4<\Delta_{2}<6$; and so on. The terms
\[
b_{k}=\frac{1}{\Delta-2k}\frac{\Gamma\left(  k+D/2\right)  }{\Gamma\left(
k+1\right)  }%
\]
in series (\ref{E6}) have the following asymptotic behavior at large $k$:%
\begin{equation}
b_{k}\approx C\,k^{D/2-2}.\label{E7}%
\end{equation}
Therefore, in the case $D\geq2$, sum (\ref{E6}) diverges if the set of basis
functions is extended ($K\rightarrow\infty$). In the limiting two-dimensional
case, series (\ref{E6}) is logarithmically divergent; at higher dimensions, it
is power-like divergent. One can verify that, in the case $D\geq2$, the
resulting energy levels of the ground and excited states become closer and
closer to the corresponding oscillator levels when $K\rightarrow\infty$; i.e.
the level shifts vanish. In particular, the ground state energy tends to the
unperturbed energy level as follows:
\begin{equation}\label{E8}
\begin{array}{l}
\displaystyle E_{0}\,\overrightarrow{_{_{_{_{K\rightarrow \infty
}}}}}\,\frac{D}{2}+\frac{\left(D-2\right)\Gamma\left(D/2\right)}
{\left(K+D/2\right)^{\left(D-2\right)/2}}, \quad D>2; \\[5mm]
\displaystyle E_{0}\,\overrightarrow{_{_{_{_{K\rightarrow \infty
}}}}}\,1+\frac{2}{\ln K+\gamma}, \quad D=2, \\
\end{array}
\end{equation}
where $\gamma=0.5772\,...$ is the Euler constant. At the same time,
the wave function tends to the unperturbed oscillator one at all
distances but the point $r=0$, where it vanishes: $\psi\left(
0\right)  =0$.

In Figure, in order to illustrate what happens to wave function
(\ref{E5}) when the basis is expanded, we show successive
approximations for the ground state wave function calculated at
various $K$'s in the three-dimensional case (for definiteness, we
took $g=1$). Similar results could be demonstrated for various
$D\geq2$ and coupling constants $g$. The larger is $K$, the smaller
is the value of the wave function at the point $r=0$, and that is
why the role of the repulsive $\delta$-potential becomes less
important (the integral of its product with the squared absolute
value of wave function tends to zero). In the limit
$K\rightarrow\infty$, the contribution of the repulsive $\delta
$-potential exactly equals zero (for $D\geq2$). Note also, that at
other points, the successive approximations tend to the unperturbed
function, though non-uniformly.

Thus, the repulsive $\delta$-potential with an arbitrary coupling constant $g$
does not shift the energy levels and does not change the wave functions of
unperturbed problem almost everywhere, except the discontinuity point at the
coordinate origin, i.e. it is not efficient in the case $D\geq2$.

It should be emphasized once again that the consideration of problem
(\ref{E3}) in the first approximation of perturbation theory with respect to
the $\delta$-potential has no sense at $D\geq2$. In particular, in the
first-order approximation for the energy levels $E_{n}^{(1)}$, one has
\begin{equation}
E_{n}^{(1)}=2n+\frac{D}{2}+g\left\vert \psi_{n}\left(  0\right)  \right\vert
^{2}, \label{E9}%
\end{equation}
whereas higher-order correction terms form a divergent series. At the same
time, the exact energy shift equals zero, in other words, the repulsive
$\delta$-potential is not efficient at $D\geq2$. This fact can be confirmed
reliably only in the framework of non-perturbative analysis. In the next
section, we give the proof of this fact on the basis of variational principle,
without use of perturbation theory.

\begin{figure}%
\vskip1mm
\includegraphics[width=\column]{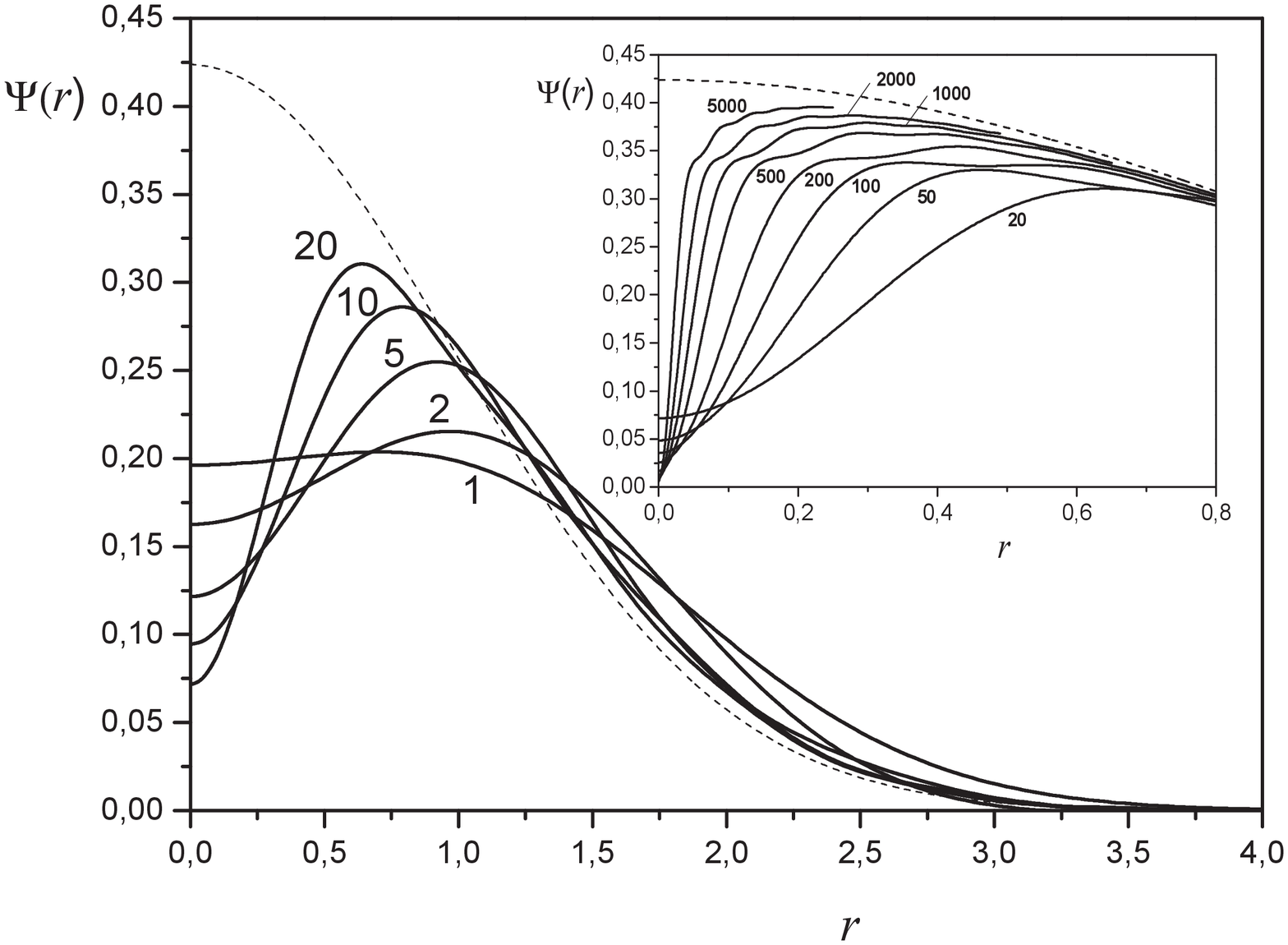}
\vskip-3mm\caption{Successive approximations of the ground-state
wave function~(\ref{E5}) of problem (\ref{E3}) in the
three-dimensional case. The numbers near the curves denote the
number $K$ of basis functions that were taken into account. In the
insert, the region of small distances is shown for large $K$'s. The
dotted line depicts the wave function of unperturbed problem
}\vspace*{2mm}
\end{figure}

To confirm the general conclusions about the role of $\delta$-like
potentials, let us demonstrate similar results obtained at $D=3$ for
another profile of external field (instead of the oscillator in
Eq.~(\ref{E3})): for a spherical rectangular well of the radius $R$,
where $V\left(  r\right)  =0$ if $r<R$, and $V\left(  r\right)
\rightarrow\infty$ if $r>R$. In addition, instead of the sequence of
functions $\delta_{\varepsilon}$ in form (\ref{E2}), we take the
potential $\delta_{\varepsilon}$ as a repulsive spherical barrier of
the radius $\varepsilon$; i.e. the potential is nonzero only at
$r<\varepsilon$, where it is constant:
$\delta_{\varepsilon}=\frac{3}{4\pi\varepsilon^{3}}g$. The passage
to the limit $\varepsilon\rightarrow0$ is done in the solutions that
are obtained in the explicit form. It can be shown directly that, at
$\varepsilon\rightarrow0$, the ground state energy of this system
approaches the value $E_{0}=\frac{\pi^{2}}{R^{2}}$ for the
unperturbed problem with the spherical potential \mbox{well as} $
E_{\varepsilon}\equiv k^{2}\,\overrightarrow{_{_{_{_{\varepsilon\rightarrow0}%
}}}}\,\frac{\pi^{2}}{\left(  R-\varepsilon\right)  ^{2}}\left(\!
1-2\sqrt
{\frac{4\pi\varepsilon}{3g}}\frac{\varepsilon}{R}+...\!\right)\! . $
The corresponding wave function at $\varepsilon<r<R$ also approaches
the unperturbed one:
$
\psi_{\varepsilon}\left(  r\right)  \,\overrightarrow{_{_{_{_{\varepsilon
\rightarrow0}}}}}\,\psi_{0}\left(  r\right)  =\frac{\sin\left(  k_{0}r\right)
}{r},
$
where $k_{0}=\pi/R$. And only within the internal region of the repulsive
potential, the wave function rapidly decreases near the coordinate origin
according to the law $\psi_{\varepsilon}\left(  r\right)  =c\,\frac
{\sinh\left(  \lambda r\right)  }{r}$, where $\lambda=\sqrt{\frac{3g}%
{4\pi\varepsilon^{3}}-k^{2}}$, being exponentially small at $r=0$: $
\psi_{\varepsilon}\left(  0\right)  =c\lambda\,\overrightarrow
{_{_{_{_{\varepsilon\rightarrow0}}}}}\,\frac{2\pi}{R}\exp\left(\!
-\sqrt {\frac{3g}{4\pi\varepsilon}}\right)\!  . $ Even in the case
$g\rightarrow\infty$ (the hard core), both the ground state energy
$E_{\varepsilon}=k^{2}=\frac{\pi^{2}}{\left(  R-\varepsilon\right)
^{2}}$ and the wave function $\psi_{\varepsilon}\left(  r\right)  =\frac{1}%
{r}\sin\left(\!  \pi\frac{\left(  R-r\right)  }{\left(
R-\varepsilon\right) }\!\right)  $ approach the corresponding
unperturbed values (in the region $r>\varepsilon$) at
$\varepsilon\rightarrow0$.\ If this conclusion is correct even for
the repulsive hard core, it is all the more correct for a weaker
repulsive potential with a finite radius and an arbitrary profile.
Note that for rapidly decreasing repulsive potentials, in
particular, like formula~(\ref{E2}), all main conclusions remain
valid irrespective of the specific potential \mbox{profile.}

Note that, in the case $D<2$, the sum in Eq.\,(\ref{E6}) converges
for both repulsive and attractive $\delta$-potentials. Hence, in
this case (in particular, in the often used one-dimensional space),
the shifts of energy levels are finite (upwards for the repulsive
$\delta$-potential), and the wave functions change substantially.
For the attractive $\delta$-potential, the ground state energy
decreases ($\Delta_{0}<0$). The same is observed for excited states,
but the corresponding shifts remain within the intervals $2\left(
n-1\right)  <\Delta_{n}<2n$. Thus, the attractive $\delta$-potential
can stimulate the appearance of only one bound state with a negative
energy in the system with Hamiltonian~(\ref{E3}) in accordance with
the fact that the $\delta$-potential is the first-rank operator.

A specific role is played by the attractive $\delta$-potentials in the case
$D\geq2$, when problem (\ref{E3}) is known to have no sense since the system
collapses. Generally speaking, in this case, the sum of a standard Hamiltonian
smoothly distributed in the space and an attractive $\delta$-potential becomes
an operator that is not bounded below, and the ground state of the system
becomes undefined. In this case, there is no sense to use expansions of type
(\ref{E5}), (\ref{E6}), because they also become undefined. At the same time,
the consideration of this problem in the first order of perturbation theory
has no sense as well, because it brings us to solution~(\ref{E9}) that has no
relation to the final result. Note also the well-known and important example:
the three-particle problem in the limit of zero force range \cite{R8}. This
limit can be interpreted as attraction between particles described by a
$\delta$-like interaction potential of type~(\ref{E2}), but the attraction
intensity also tends to zero at $\varepsilon\rightarrow0$ following a law that
allows a finite (given a priori) energy of the two-particle bound state to be
fixed. However, even in the case of such a \textquotedblleft
weaker\textquotedblright\ $\delta$-like attraction potential,\textrm{\ }the
system of three and more particles will collapse. The collapse in the system
of three particles in the three-dimensional case and in the limit of zero
force range was studied in works \cite{R9, R10} in detail. The cited authors
even found the law describing how the energy levels, the number of which
becomes infinitely large, tend to minus infinity and predicted the main
relations for the phenomenon, which is now known as the Efimov effect.

Conclusions drawn for problem~(\ref{E3}), irrespective of whether the $\delta
$-potential is repulsive or attractive, are completely confirmed by other
examples of similar problems. In particular, besides the example of
interaction in an external area in the form of a spherical potential well,
which was mentioned above, we could take other exactly solvable problems,
e.g., a $D$-dimensional cubic box with infinitely high walls or the Coulomb
potential. In all those cases, the divergence of (\ref{E6})-type series at
$D\geq2$ leads to results similar to Eq.~(\ref{E8}). The main conclusion
concerning the inefficiency of repulsive $\delta$-potential in the case
$D\geq2$ remains valid. Note that, for problems with external potentials that
generate a finite number of discrete levels and a continuous spectrum, while
considering a generalized expression of type (\ref{E6}) with an additional
(besides the summation over the discrete spectrum) integration over the
continuum spectrum, all the principal conclusions concerning the inefficiency
of $\delta$-potential remain in force.

In the next section, we prove the statements made above with the help of
variational principle used in the framework of rather general assumptions
concerning the Hamiltonian that contains a repulsion in the form of $\delta$-potential.

\section{Proof on the Basis of Variational Principle}

First, let us consider in detail the case $D=2$, which is the most delicate
for proof. This is a \textquotedblleft critical\textquotedblright\ case,
because, the repulsive $\delta$-potential becomes efficient at $D<2$: it
affects the physical observable quantities and changes the wave functions. We
accept the most general assumptions for the system Hamiltonian unperturbed by
the $\delta$-potential: the only requirement is that the wave functions of the
unperturbed problem should be finite at $r=0$, which is reasonable for a wide
class of commonly used potentials $V\left(  r\right)  $. Let the ground state
of the system have the energy $E_{0}$ and be described by the wave function
$\psi_{0}\left(  r\right)  $, with the very existence of the ground state
being determined by the potential $V\left(  r\right)  $. In order to
variationally estimate the ground state energy, keeping in mind the
preliminary consideration of the possible influence of $\delta$-potential on
the system and understanding that the account of the wave function behavior at
small distances is crucially important, we construct the trial wave function
$\psi\left(  r\right)  $ in the form
\begin{equation}
\psi\left(  r\right)  =f\left(  r\right)  \psi_{0}\left(  r\right)
.\label{E10}%
\end{equation}
The correlation factor is chosen to equal $f\left(  r\right)  \equiv
1-\exp\left(\!  -\!\left(  \beta r\right)  ^{1/\alpha}\!\right) $,
where the parameter $\alpha=\alpha\left(  \beta\right)  $ is an
infinitely increasing function of $\beta$, i.e. $\alpha\left(
\beta\right) \,\overrightarrow
{_{_{_{_{\beta\rightarrow\infty}}}}}\,\infty$. Omitting the general
analysis of allowable functions $\alpha=\alpha\left(  \beta\right)
$, we restrict ourselves to the example $\alpha\equiv\ln\left(
\ln\beta\right)  $, which is sufficient for the proof of our
statement. The energy of the ground state is denoted by $E_{0}$, and
the corresponding wave function of the unperturbed (by the
$\delta$-potential) problem by $\psi_{0}\left(  r\right)  $. The
variational estimate for the energy of the ground state of the
Hamiltonian with an additional repulsive $\delta$-potential made
using wave functions
(\ref{E10}) can be presented in the form%
\[
E\leq\frac{\int \psi^{\ast}\left( r\right)
\left(-\frac{1}{2}\triangle+V\left(r\right)+g\delta\left(\mathbf{r}
\right)\right)\psi\left(r\right)d\mathbf{r}} {\int
\left|\psi\left(r\right)\right|^{2}d\mathbf{r} }\equiv
\]\vspace*{-5mm}
\begin{equation}\label{E11}
\equiv E_{0}+ \frac{\frac{1}{2}\int
\left|\psi_{0}\left(r\right)\right|^{2}\left(\nabla
f\left(r\right)\right)^{2}d\mathbf{r}}{\int
\left|\psi_{0}\left(r\right)\right|^{2}
f^{2}\left(r\right)d\mathbf{r} },
\end{equation}
where the integration is carried out over the two-dimentional space. Note that
the $\delta$-potential itself makes no contribution to the numerator in
Eq.~(\ref{E11}) because of the correlation factor property $f\left(  0\right)
=0$. However, owing to the same correlation factor, there arises another,
additional to $E_{0}$, term in Eq.~(\ref{E10}), which follows from the kinetic
energy operator. If the normalized wave function of unperturbed problem is
finite, i.e. $\left\vert \psi_{0}\left(  r\right)  \right\vert ^{2}\leq C_{0}$
(which is valid for an arbitrary non-singular Hamiltonian), the normalization
integral in the denominator of Eq.~(\ref{E11}) tends to unity at
$\beta\rightarrow\infty$,
\begin{equation}\label{E12}
\int\left|\psi\left(r\right)\right|^{2}d\mathbf{r}
\,\overrightarrow{_{_{_{_{\beta\rightarrow \infty }}}}}\,
1+\mathcal{O}\left(\!\frac{\alpha
\Gamma\left(2\alpha\right)}{\beta^{2}}\!\right)
\,\overrightarrow{_{_{_{_{\beta\rightarrow \infty }}}}}\,1,
\end{equation}
since $\alpha=\ln\left(  \ln\beta\right)  $ and due to the
asymptotic properties of gamma function $\Gamma\left(  z\right)  $.
Let us consider now the integral in the numerator of
Eq.~(\ref{E10}),
\[
\frac{1}{2}\int\left|\psi_{0}\left(r\right)\right|^{2}\left(\nabla
f\left(r\right)\right)^{2}d\mathbf{r}\leq \frac{1}{2}C_{0}
\int\left(\nabla f\left(r\right)\right)^{2}d\mathbf{r}=
\]\vspace*{-7mm}
\begin{equation}\label{E13}
=\frac{\pi C_{0}}{4\alpha}.
\end{equation}
Hence, we have $E\leq E_{0}+\frac{\pi
C_{0}}{4\alpha}\,\overrightarrow
{_{_{_{_{\beta\rightarrow\infty}}}}}\,E_{0}$. Since the repulsive
$\delta $-potential can shift the energy only upwards, i.e. $E\geq
E_{0}$, we ultimately obtain that $E=E_{0}$ in the two-dimensional
case.

The proof becomes essentially simpler for spaces of higher dimensionalities,
because the larger is $D$, the more important role plays the $r^{D-1}$ factor
in integration at small distances. Already for $D>2$, it is sufficient to
choose a simpler correlation factor in trial function (\ref{E10}), e.g.,%
\begin{equation}
f\left(  r\right)  =1-\exp\left(  -\left(  r/b\right)  ^{2}\right)\!
,
\label{E14}%
\end{equation}
and pass to the limit $b\rightarrow0$ in final expressions. Using correlation
factor (\ref{E14}), let us consider the problem of $\delta$-potential
efficiency for the systems of $N$ particles in the case $D>2$. To carry out
the variational estimation of the ground state energy, we use the trial
variational function in the form
\begin{equation}
\Psi=\Psi_{0}F\left(  r_{12},r_{13},...\right)  =\Psi_{0}\prod
\limits_{i>j=1}^{N}f\left(  r_{ij}\right)\!  , \label{E15}%
\end{equation}
where the pair correlation factors $f\left(  r_{ij}\right)  $ have
form (\ref{E14}), and $\Psi_{0}$ is the wave function of the ground
state for the Hamiltonian $H_{0}$ that differs from
Hamiltonian~(\ref{E1}) by ~the~ absence~ of
the term $\delta V= $ $=g\sum_{i>j=1}^{N}\delta\left(  \mathbf{r}%
_{ij}\right)  $. For the ground state energy of problem (\ref{E1}), we obtain
the following variational estimate (similar to Eq.~(\ref{E11})):
\[
E \leq
E_{0}+\frac{\frac{1}{2m}\sum\limits_{i=1}^{N}\int\left(d\mathbf{r}\right)^{N}
\left|\Psi_{0}\right|^{2}\left(\left(\nabla_{i}F\right)^{2}+\delta V
F^{2} \right)} {\int\left(d\mathbf{r}\right)^{N}
\left|\Psi_{0}\right|^{2}F^{2}}=
\]\vspace*{-7mm}
\begin{equation}\label{E16}
=E_{0}+\frac{\frac{1}{2m}\sum
\limits_{i=1}^{N}\int\left(d\mathbf{r}\right)^{N}
\left|\Psi_{0}\right|^{2}\left(\nabla_{i}F\right)^{2}}
{\int\left(d\mathbf{r}\right)^{N} \left|\Psi_{0}\right|^{2}F^{2}}.
\end{equation}
The potential $\delta V$ disappears from the numerator of Eq.~(\ref{E16}) due
to the following correlation factor property: $F=0$ if any of the pair
distances $r_{ij}=0$. We emphasize once more that the term additional to
$E_{0}$ in Eq.~(\ref{E16}) originates from the kinetic energy operator action
on the correlation factors. We omit the straightforward but cumbersome
calculations of the derivatives of $F$, as well as the estimation of integrals
in Eq.~(\ref{E16}) at $b\rightarrow0$, and give the ultimate estimate for the
energy (under the natural assumption that the squared wave function is finite,
$\left\vert \Psi_{0}\right\vert ^{2}\leq C_{0}$):
\begin{equation}
E\leq E_{0}+\mathcal{O}\left(  b^{D-2}\right)\!  . \label{E17}%
\end{equation}
The higher is the dimension $D$ of the space, the more rapidly tends
this value to the unperturbed value $E\leq$ $\leq E_{0}$ as
$b\rightarrow0$. It is clear a priori that any repulsive potential
$\delta V$ can result only in $E\geq E_{0}$. Therefore, we
ultimately obtain that $E=E_{0}$.

We point out that, in the case $D=2$, the correlation factor may be taken in
the form used in Eq.~(\ref{E10}) for the two-particle problem, and the above
mentioned arguments can be repeated for the system of $N$ particles. But
instead of formula~(\ref{E17}), we will obtain an estimation of type
(\ref{E13}).

Note that the obtained results concerning the inefficiency of a repulsive
$\delta$-potential in spaces with $D\geq2$ dimensions can be directly
generalized to the excited energy levels. To this end, it is sufficient to
take into account that the trial variational wave functions of the $n$-th
excited state should be orthogonal to the functions corresponding to the lower
levels. This requirement is satisfied automatically for the functions of type
(\ref{E10}) or (\ref{E15}) (where $\Psi_{0}$ is to be substituted by $\Psi
_{n}$) in the limit of zero correlation radius, due to the orthogonality of
the wave functions in the unperturbed problem.

It is important to understand that the repulsive $\delta$-potential does not
affect other physical observable quantities. In particular, it was shown in
work~\cite{R11} that the repulsive $\delta$-potential does not change the
phase shifts in two-dimensional case, which is the most delicate for the
proof. This means that, if $D>2$, the obtained result is even more valid
(unfortunately, this fact was not emphasized in work~\cite{R11}). In any case,
if a repulsive $\delta$-potential affects neither the spectrum nor the phase
shifts, it is inefficient at $D\geq2$.

\section{Conclusions}

To summarize, it was shown that a repulsive $\delta$-potential is inefficient
for a quantum system of particles with an interaction containing $\delta
$-potentials if the space dimensionality $D\geq2$. The consideration of such
problems in the first order of perturbation theory leads to incorrect results.
The consistent account of short-range correlations demonstrates that such
potentials make no influence on the energy spectrum and other physical
observable quantities. On the other hand, in the case $D\geq2$, attractive
$\delta$-potentials are known to produce a system collapse. So, the $\delta
$-potentials are efficient only for one-dimensional problems. But for a system
in a $D$-dimensional space when $D\geq2$ and in the framework of the accurate
problem formulation, there is no sense to use such interaction potentials at all.

\vskip3mm{\it The authors thank Prof.\,O.S.\,Bakai,
Prof.\,I.M.\,Bur\-ban, and Prof. P.M.\,Tomchuk for useful discussion
of some issues concerned.

This work was partially supported by the Program of Fundamental
Research of the Department of Physics and Astronomy of the National
Academy of Sciences of Ukraine (project No.~0112U000056).}

\rezume{%
І.В.\,Сименог, Б.Є.\,Гринюк, М.В.\,Кузьменко}{ПРО НЕЕФЕКТИВНІСТЬ\\
ВІДШТОВХУВАЛЬНИХ $\delta$-ПОТЕНЦІАЛІВ\\ У БАГАТОВИМІРНИХ ПРОСТОРАХ}
{Показано, що $\delta$-подібні відштовхувальні потенціали взаємодії
при повному врахуванні кореляцій є неефективними для будь-якої
$N$-частинкової квантової системи у $D$-вимірному просторі при $D
\geq 2$.}


\begin{thebibliography}{99}                                                                                               %


\bibitem {R1}A.I.~Akhiezer and V.B.~Berestetskii, \textit{Quantum
Electrodynamics} (Interscience, New York, 1965).

\bibitem {R2}E.P.~Gross, Nuovo Cimento. \textbf{20}, 454 (1961).

\bibitem {R3}L.P.~Pitaevskii, Zh. \`{E}ksp. Teor. Fiz. \textbf{40}, 646 (1961).

\bibitem {R4}B.B.~Kadomtsev and M.B.~Kadomtsev, Usp. Fiz. Nauk \textbf{167},
649 (1997).

\bibitem {R5}T.H.R.~Skyrme. Nucl. Phys. \textbf{9}, 615 (1959).

\bibitem {R6}D.~Vautherin and D.M.~Brink. Phys. Lett.~B \textbf{32}, 149 (1970).

\bibitem {R7}P.~Ring and P.~Schuck, \textit{The Nuclear Many-Body Problem}
(Springer Verlag, New York, 1980).

\bibitem {R8}G.V.~Skornyakov and K.A.~Ter-Martirosyan, Zh. \`{E}ksp. Teor.
Fiz. \textbf{31}, 775 (1956).

\bibitem {R9}R.A.~Minlos and L.D.~Faddeev, Dokl. Akad. Nauk SSSR \textbf{141},
1335 (1961).

\bibitem {R10}R.A.~Minlos and L.D.~Faddeev, Zh. \`{E}ksp. Teor. Fiz.
\textbf{41}, 1850 (1961).

\bibitem {R11}V.V. Babikov, \textit{Phase Function Method in Quantum
Mechanics} (Nauka, Moscow, 1988) (in Russian).

\begin{flushright}
{\footnotesize Received 24.09.2014.\\ Translated from Ukrainian by
O.I.~Voitenko}
\end{flushright}
\end{thebibliography}
\end{document}